# Label-free, non-contact, *in-vivo* ophthalmic imaging using photoacoustic remote sensing microscopy

ZOHREH HOSSEINAEE[1], LAYLA KHALILI[1], JAMES ALEX TUMMON SIMMONS[1], KEVAN BELL[1,2], PARSIN HAJI REZA[1*]

[1]*PhotoMedicine Labs, Department of System Design Engineering, University of Waterloo, Ontario, N2L 3G1, Canada*
[2]*illumiSonics, Inc., Department of Systems Design Engineering, University of Waterloo, Waterloo, Ontario N2L 3G1, Canada*
*\* phajireza@uwaterloo.ca*

**Abstract:** We present the first label-free, non-contact, *in-vivo* imaging of the ocular vasculature using photoacoustic remote sensing (PARS) microscopy. Both anterior and posterior segments mouse eye were imaged. Vasculature of iris, sclera and retina tissues were clearly resolved. To best of our knowledge this the first study showing non-contact photoacoustic imaging conducted on *in-vivo* ocular tissue. We believe that PARS microscopy has the potential to advance the diagnosis and treatment of ocular diseases.

## 1. Introduction

Ophthalmic imaging has long played an important role in the understanding, diagnostic and treatment of a wide variety of ocular disorders. Currently available clinical ophthalmic imaging instruments are primarily optical-based, including slit-lamp microscopy, fundus photography, confocal microscopy, scanning laser ophthalmoscopy and optical coherence tomography (OCT). Despite offering valuable structural and morphological information from ocular tissue, these modalities have limitations providing detailed functional and molecular characteristics of the eye[1]. Access to these information would help with early diagnosis and consequently facilitate the treatment of major eye diseases including age related macular degeneration, glaucoma and diabetic retinopathy[2].

Photoacoustic microscopy (PAM) is among the most rapidly growing optical imaging modalities. The technology is well-known for its practical functional and molecular imaging capabilities. Its unique imaging contrast of optical absorption makes PAM the preferred modality for a wide range of biomedical applications[3]. In ophthalmic imaging, PAM has been used for visualizing hemoglobin and melanin content[4], measuring blood oxygen saturation[5], and quantifying metabolic rate of oxygen in the ocular tissue[6,7]. Even though PAM offers high sensitivity, unique imaging contrast, and high resolution, it is not generally an all-optical imaging method unlike the other ophthalmic microscopy techniques. One of the significant limitations of photoacoustic microscopes arises from their need to be in physical contact with the sample through a coupling media. This physical contact, coupling, or immersion of the sample is undesirable in ophthalmic applications. It may increase the risk of abrasion, infection, and patient discomfort. Additionally, involuntary eye movements may affect the coupling efficiency and degrade image quality. In small animal imaging, immersion in water significantly complicates the procedure and commonly results in sacrificing the animal[8,9].

To overcome the limitations of contact-based PAM devices, in 2017 Haji Reza et. al developed photoacoustic remote sensing (PARS) microscopy for non-contact, non-interferometric detection of photoacoustic signals[10]. The technology has proved its potential over a short period of time in various biomedical applications such as label-free histology imaging[11,12], $SO_2$ mapping and angiogenesis imaging[13]. For a long time it has been a desire to achieve non-contact photoacoustic ophthalmic imaging[14]. Here, we have successfully extended the application of PARS microscopy for non-contact photoacoustic imaging of ocular tissue. To make this possible, we made significant modifications to make PARS microscopy suitable for ophthalmic imaging. These modifications include, eye-friendly detection wavelength, suitable scanning pattern and interpolation algorithm, and focusing optics. For example, compared to conventional PARS systems that employ 1310 nm detection beam, here 830 nm detection beam is used. The novel 830 nm detection improves photoacoustic signal detection in the ocular environment by having lower absorption in water. Additionally, it reduces the amount of chromatic aberration in the system by having close spectral bandwidth to the 532 nm excitation beam. The performance of the system is demonstrated for imaging different regions of the ocular tissue including, iris, scleral and retinal vasculature. To best of our knowledge this is the first study showing non-contact photoacoustic imaging conducted on ocular tissue.

## 2. Methods

### 2.1 Optical design

Figure 1 illustrates the experimental setup used in this study. The excitation source is 532-nm 1 ns pulse-width, ytterbium-doped fiber laser (IPG Photonics) capable of pulse repetition rates from 20 kHz to 600 kHz. The detection arm uses an 830-nm Superluminescent Diodes with 20 nm full width at half maximum linewidth (SLD830S-A20, Thorlabs). The output end of the fiber is coupled to a collimator. A polarized beam splitter is used to transmit the majority of the forward light onto a quarter wave-plate, which transforms the linearly polarized light to circularly polarized light. The detection and excitation light are then combined using a dichroic mirror. The co-aligned beams are then directed toward a large beam galvanometer scanning mirror system (GVS012/M, Thorlabs, Inc.) driven by a two-channel function generator. The beams are then directed to a set of 1:1 telecentric pair in order to provide uniform image intensity and improve the effective imaging field of view. A 0.26 NA reflective objective co-focused the beams onto the sample. The back-reflected light from the sample is collected via the same objective lens and guided towards the detection path. The quarter wave-plate transforms the reflected circularly polarized light back to linearly polarized light. This enables the polarized beam splitter to direct the back-reflected light towards the photodiode. A long-pass filter (FELH0800, Thorlabs Inc.) is used to block any residual 532 nm light. The 830 nm signal is then focused onto the photodiode with an aspherical lens. The photodiode is connected to a high-speed digitizer (CSE1442, Gage Applied, Lockport, IL, USA). A point acquisition is acquired for each pixel and recorded by the digitizer. Each point acquisition is converted to an intensity value by computing its maximum amplitude and plotted at its respective location in the image.

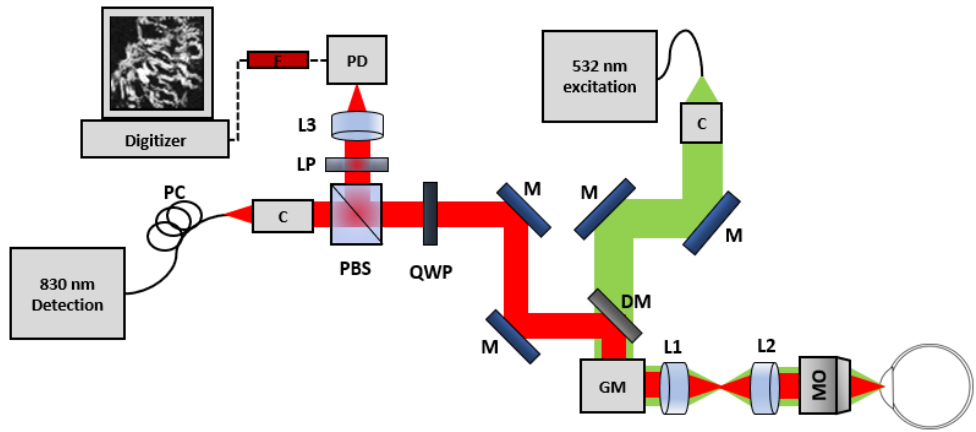

**Figure 1.** Schematic of the PARS microscopy system. M: Mirror, F: digital filter, PC: polarization

controller, DM: Dichroic mirror, QWP: Quarter wave plate, PBS: Polarized beamsplitter, LP: Long pass filter, GM: Galvanometer mirrors, MO: microscope objective, L: Lens, C: Collimator, PD: Photodiode.

## 3. Results

The performance of the system was first tested on 7 µm carbon fiber phantom and representative images are shown in Figure 2 A-C. The images were acquired using ~ 900 pJ excitation pulse energy and ~ 2 mW interrogation power on the sample. The signal-to-noise ratio (SNR), defined as the average of the maximum amplitude projection pixels in a region of interest over the standard deviation of the noise, was quantified as 56±3 dB.

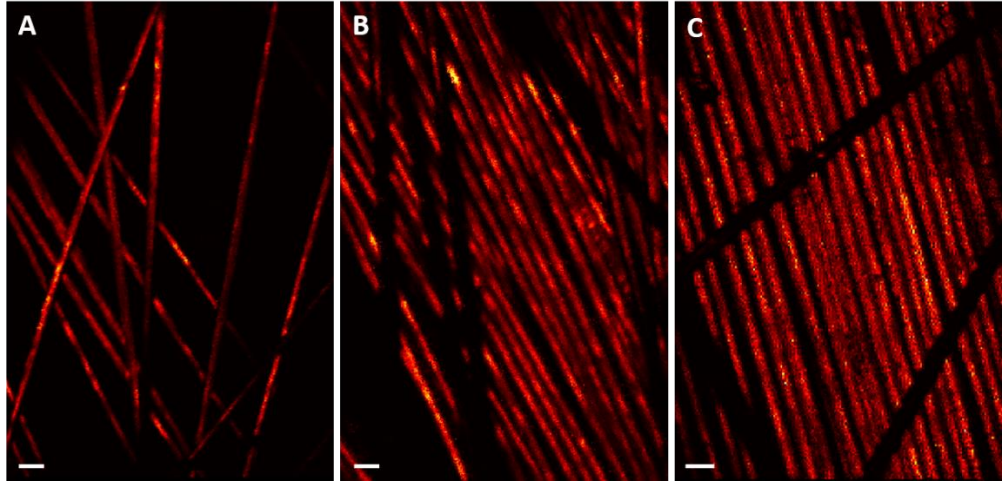

**Figure 2**. Carbon fiber images acquired using PARS microscopy. Scale bar: 30 µm.

To demonstrate the *in-vivo* capabilities of the system the ear of a nude mouse (NU/NU, Charles River, MA, USA) was imaged. All of the experimental procedures were carried out in conformity with the laboratory animal protocol and was approved by the Research Ethics Committee at the University of Waterloo. A custom-made animal holder was used to restrain the animal. The base of the animal holder was lined with a thermal pad in order to keep the mouse body temperature between 36° and 38°C. Artificial tears were used frequently (~ every 5 minutes) to keep the cornea hydrated. Vital signs, such as respiration rates, heart rates and body temperature were monitored during the experiment. All of the 2D images shown in this manuscript were formed using a maximum amplitude projection (MAP) of each A-scan as a pixel in a C-scan en-face image. All images shown in this manuscript were produced by direct plotting from interpolated raw data using a Delaunay triangulation interpolation algorithm[15]. All images and signal processing steps were performed in the MATLAB environment. Frangi vesselness filter was applied on the vasculature images[16]. Scale bars in the field of view (FOV) were calibrated using a 1951 USAF resolution test target.

Figure 3 demonstrates *in-vivo* PARS images of en-face microvasculature in the ear. Images were acquired with 20 KHz pulse repetition rate (PRR) of the excitation laser. The field of view covered in Figure 3A is 1 mm × 1 mm, and it took ~7 seconds to acquire the image. Figure 3B, was recorded from the same area with a smaller field of view of 500 µm × 500 µm, red blood cells within the capillaries can be clearly seen in this image. The lateral resolution of the system for *in-vivo* experiment was calculated ~ 2.6 µm. The measured pulse energy at the sample surface was measured as ~ 50 nJ and the detection power was ~4 mW. The SNR of the large vessels was measured as approximately 41± 4 dB.

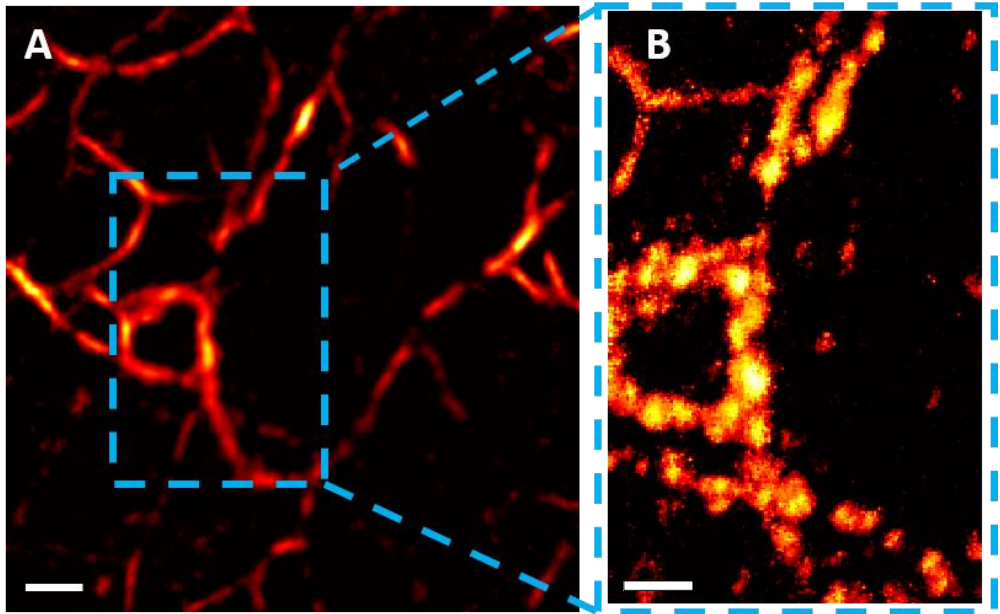

**Figure 3.** In-vivo imaging of mouse ear using photoacoustic remote sensing microscopy (A) vasculature of the ear in a 1 mm × 1 mm area, Scale bar: 100 µm. (B) Ear vasculature recorded from smaller field of view in a 500 µm × 500 µm area. Scale bar: 15 µm.

For the eye imaging experiments, the repetition rate of the excitation laser was increased to 100 KHz to reduce the effect of motion artifacts. Both iris and retinal vasculature were imaged *in-vivo* in the mouse eye. Figure 4 depicts representative images acquired from the iris vasculature within different field of views. Figure 4A represents an image acquired from the vasculature in the inferior iris, covering an area of 2.5 mm × 1 mm, and it took ~ 3 seconds to record the image. The enlarged capillary networks of this region are shown in Figure 4B & C. Figure 4D is recorded from the vasculature of the peripheral iris, and Figure 4E shows vascular network near the limbal and episcleral region in the mouse eye. Media 1. demonstrates Live feedback of vasculature in this region acquired during manual depth scanning. To maintain higher real-time speed, a more basic scatter point interpolation was used, which resulted in lower resolution compared to single captures. The live feedback was used constantly during the imaging session to enable accurate alignment and focusing. Unlike other pre-clinical imaging techniques[17,18], in these experiments (to mimic real world situation for clinical applications) the head of animal was not fixed and no tropical anesthesia was applied to the eyeball. Therefore, motion artifacts are still presented in the images.

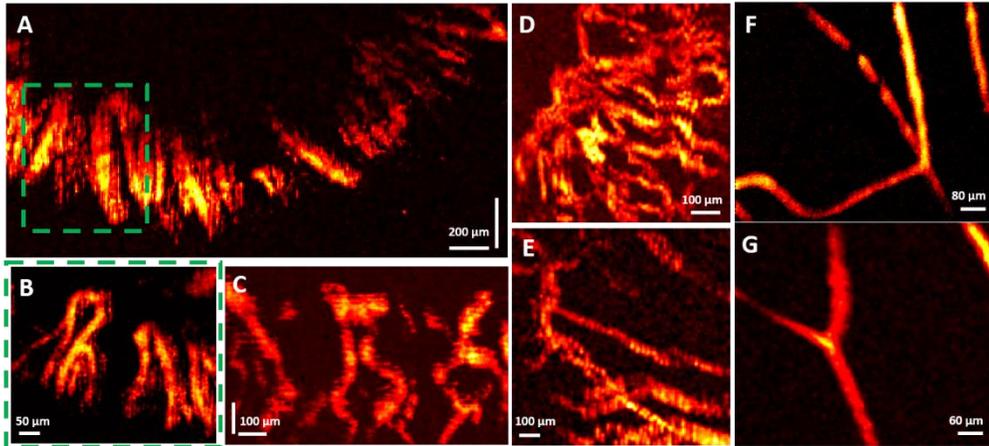

**Figure 4.** Iris vasculature within different field of views. (A) Vasculature in the inferior iris, covering an area of 2.5 mm × 1 mm. (B & C) The enlarged capillary networks of inferior iris. (D) Peripheral iris vasculature. (E) Vascular network near the limbal and episcleral region in the mouse eye. (F) Branches of retinal vasculature network in 600 µm × 600 µm FOV. (G) Branches of retinal vasculature in a 450 µm × 450 µm FOV.

Since the employed objective lens of the system has long enough working distance (~ 30 mm) compared to the diameter of mouse eyeball (~3 mm), imaging the retinal vasculature was also possible using the current setup. Figure 4F & G show representative images acquired from retinal vasculature. The SNR of the large vessels was measured as approximately 37 ± 3 dB. Both detection and excitation beams were sent through the central part of the anterior segment to back of the eye. The reflected signals coming from the retinal vasculature were detected to form the images. The images could be formed with minimum pulse energy of ~ 50 nJ. Based on previously reported values (40 nJ - 80 nJ)[19] for photoacoustic eye imaging the pulse energy used in this study is in the safe eye pulse energy range, however, the system still needs to be improved to be safe for clinical trials. The ~ 4 mW detection power is also within the ANSI safety limits[20].

There are several aspects that can be further refined for future studies. The field of view of the current setup is limited to 2.5 mm × 2.5 mm, which is not enough for capturing the full iris/retina vasculature network in rodent eye. To improve the field of view we plan to employ customized scan lens. Moreover, we plan to apply multi-wavelength PARS to assess the blood oxygen saturation in both anterior and posterior segments of the eye in rodent models. We can potentially integrate optical coherence tomography to image with both scattering and absorption contrast and acquire volumetric images of the eye structure and vasculature. Since PARS is an all-optical imaging system, it can naturally be combined with all other all-optical imaging systems such as OCT. In addition, the PARS detection laser can be modified to act as an OCT beam to facilitate this integration.

In summary PARS microscopy was used to conduct non-contact, label-free, *in-vivo* photoacoustic imaging of ocular tissue for the first time. Successful imaging of the ocular vasculatures in the mouse eye demonstrate the capability of PARS microscopy for *in-vivo* ocular imaging. We believe the presented results are major steps towards introducing photoacoustic imaging systems to the clinical ophthalmic setting. Our current study laid the foundation for future quantitative imaging of the $sO_2$ in the ocular vasculature and also measuring the distribution of the pigments in the retinal pigment epithelium (RPE) layer, which are important for the research and clinical diagnosis of ocular diseases.

**Conflict of Interest**
Authors K. Bell and P. Haji Reza have financial interests in illumiSonics Inc. IllumiSonics partially supported this work.

**Funding**
New Frontiers in Research Fund – Exploration (NFRFE-2019-01012); Natural Sciences and Engineering Research Council of Canada (DGECR-2019-00143, RGPIN2019-06134); Canada Foundation for Innovation (JELF #38000); Mitacs (IT13594); Centre for Bioengineering and Biotechnology (CBB Seed fund); University of Waterloo; illumiSonics (SRA #083181).

**Acknowledgments**
The authors would like to thank Jean Flanagan for assistance with the animal-related procedures. The authors would also like to thank Ben Ecclestone, Nicholas Pellegrino and Marian Boktor for their support and help. The authors acknowledge funding from the University of Waterloo, NSERC Discovery grant, MITACS accelerator program, Canada Foundation for Innovation (CFI-JEFL), Centre for Bioengineering and Biotechnology seed funding, New Frontiers in Research Fund –exploration, and research partnership support from illumiSonics Inc.